\documentclass[twocolumn,prb,12point]{revtex4-1}

\usepackage{amsmath, amsfonts, amssymb}
\usepackage{graphicx}

\begin{document}

\title{The evolution of oscillator wave functions}
\author{Mark Andrews}
\email{mark.andrews@anu.edu.au}

\affiliation{Department of Quantum Science, Australian National University, ACT2601, Australia}

\date{\today}

\begin{abstract}
We consider some of the methods that can be used to reveal the general features of how wave functions evolve with time in the harmonic oscillator. We first review the periodicity properties over each multiple of a quarter of the classical period of oscillation. Then we show that any wave function can be simply transformed so that its centroid, defined by the expectation values of position and momentum, remains at rest at the center of the oscillator. This implies that we need only consider the evolution of this restricted class of wave functions; the evolution of all others can be reduced to these. The evolution of the spread in position $\Delta_x$ and momentum $\Delta_p$ throws light on energy and uncertainty and on squeezed and coherent states. Finally we show that any wave function can be transformed so that $\Delta_x$ and $\Delta_p$ do not change with time and that the evolution of all wave functions can easily be found from the evolution of those at rest at the origin with unchanging $\Delta_x$ and $\Delta_p$.
\end{abstract}


\maketitle 

\section{Introduction}
\label{sec:intro}
The simple harmonic oscillator is studied in every course in elementary quantum mechanics because of its relative mathematical simplicity and its usefulness as a model of natural processes: oscillators are ubiquitous in nature and harmonic traps are used in many labs. Yet little attention is given to the time-evolution of oscillator wave functions, beyond the fact that the evolution is periodic. Many texts do derive the propagator, but using it requires an integration that can be carried out exactly only for a small class of wave functions.\cite{Anal}

Here we will consider some of the simple methods that can be used to understand the general features of how oscillator wave functions evolve. We start with the simple relations between the evolved and the initial wave functions after each multiple of $\frac{1}{4}T$, where $T$ is the period of the equivalent classical oscillator. Then the evolution of some moments of the wave functions will be determined. The first moments, that is the expectation values of position and momentum, give Ehrenfest's result that $\langle\hat x\rangle$ and $\langle\hat p\rangle$ follow a classical trajectory. A powerful result is that the evolution of oscillator wave functions is essentially independent of the first moments. Consequently we need consider only states that remain centered on the origin (taken to be the point where the oscillator restoring force is zero). All other states can be reduced to these by a simple transformation. 

The second moments give a rich set of results about uncertainties and energy. Squeezed states emerge as Gaussian states at the extreme edge of a strong form of the uncertainty relations, an inequality involving the second moments. All Gaussian states can be produced by changing the spatial scale of (that is squeezing) the ground state, and applying the transformations discussed in the previous paragraph. Similarly, any state can be produced, apart from a phase, by changing the scale of its `stable' form whose second moments remain constant as it evolves. The evolution of any state can be obtained from that of its stable form.

A note on the use of dimensionless variables: the oscillator has its own natural length scale $\alpha=(\hbar/m\omega)^{1/2}$ and the theory becomes much simpler in terms of the dimensionless variables $x/\alpha$, $\alpha p/\hbar$ and $\omega t$. The oscillator has a symmetry between position and momentum (essentially because each appears only as a quadratic term in the Hamiltonian) and this is somewhat obscured when using the usual physical quantities $x$ and $p$. Almost every equation in this paper would be simpler, and the general arguments clearer, if expressed in the dimensionless variables. The downside would be that it is harder for students to relate the oscillator theory to that of other systems if the discussion is in terms of special oscillator variables. For that reason physical variables will be used, except for Eq.~(\ref{eqn:FT}) of Section \ref{sec:prop} and all of Appendix B, which involve Fourier transforms.

\section{Periodicity}
\label{sec:Period}
For the oscillator with mass $m$ and angular frequency $\omega$, the Hamiltonian is
\begin{equation}\label{eqn:ham}
\hat{H} = \frac{1}{2m}\hat{p}^2 + \frac{1}{2}m\omega^2 \hat x ^2
\end{equation}
and the energy levels, that is the eigenvalues of $\hat H$, are $E_n = \hbar \omega (n+\frac{1}{2})$ with $n = 0, 1, 2, ...$. If a wave function $\psi (x)$ is expressed as a sum over the energy eigenfunctions $\psi_n (x)$ at time $t=0$, so that $\psi (x)=\sum_{n=0}^{\infty} c_{n}\psi_n (x)$, then the evolved wave function at time $t$ is 
\begin{equation} \label{eqn:eexpn}
\psi(x,t)=\sum_{n=0}^{\infty}\exp[-\imath \omega t(n+\small{\frac{1}{2}})]c_{n}\psi_n (x),
\end{equation}
where $\imath=\sqrt{-1}$ and the $c_n$ are generally complex.

\vspace{1.5mm}
\hspace{-3.5mm}\textbf{Evolution over one period}. If $t=2\pi/\omega$, the factor $\exp[-\imath \omega t(n+\frac{1}{2})]$ becomes $-1$, so that $\psi(x,T)=-\psi(x,0)$ where $T=2\pi/\omega$. Thus, every wave function returns exactly to its original form after each period $T$, except that the wave function changes sign. This period $T$ is also the period of the corresponding classical oscillator. This periodicity is a direct consequence of the equal spacing of the energy eigenvalues.

\vspace{1.5mm}
\hspace{-3.5mm}\textbf{Evolution over half a period}. At time $t=\pi/\omega$, $\exp[-\imath \omega t(n+\frac{1}{2})]$ becomes $-\imath(-1)^n$. Since the energy eigenfunctions are alternately even and odd, $\psi_n(-x)=(-1)^n\psi_n(x)$, it follows from Eq.~(\ref{eqn:eexpn}) that
\begin{equation}
\psi(x,T/2)=-\imath\psi(-x,0).
\end{equation}
This holds for any initial time and therefore
\begin{equation}
\psi(x,t+T/2)=-\imath\psi(-x,t).
\end{equation}
\textbf{Evolution through the second quarter period}. For an initially real wave function, the evolution for all time can be deduced from that over the first quarter period. Inserting time $t'=T/2-t$ into Eq.~(\ref{eqn:eexpn}) leads to $\psi(x,t')=-\imath\sum (-1)^n c_n\exp[\imath\omega t(n+\frac{1}{2})]\psi_n(x)$. Since, $\psi_n(-x)=(-1)^n\psi_n(x)$, and the $c_n$ are real, we have
\begin{equation}\label{eqn:q2}
\psi(x,T/2-t)=-\imath\psi^\star(-x,t)\hspace{5mm}\textrm{if}\:\psi(x,0)\:\textrm{real}.
\end{equation}
Thus, after one quarter period, $|\psi(x,t)|$ retraces the values it took during the first quarter, but it is spatially reversed  and the real part takes the negative of the values that the imaginary part took. Only the evolution over one quarter period need be calculated and the rest can be deduced. An example is shown in Fig.~1 with more detail in Section IX.

\begin{figure}[h!]
\centering
\includegraphics{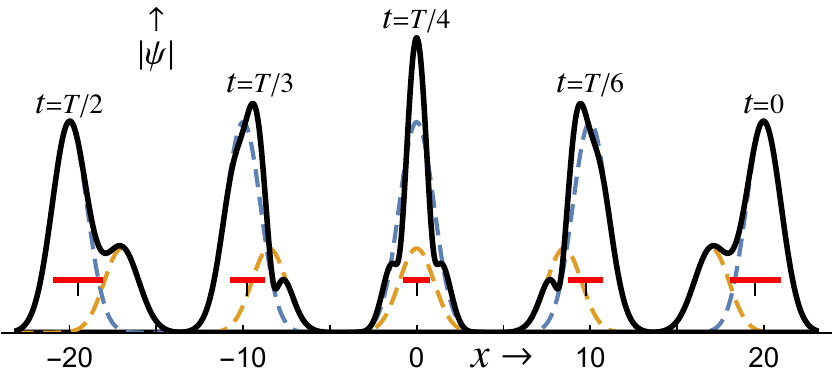}
\caption{\label{dub}}
The evolution over a half period of a superposition $\psi$ of two  Gaussian wave packets, each with the same width as the ground state. The larger packet starts at rest with its center at $x=20\alpha$ while the smaller, with amplitude $0.4$ of that of the larger, starts at $17\alpha$ and also at rest. The solid curves show $|\psi|$ for five times over a half period while the dashed curves show the magnitude of each Gaussian if the other were not present. The thick horizontal lines show the spread in position $\Delta_x$ of the superposition. The initial wave function is real and therefore the graphs for negative $x$ are the mirror image those for positive $x$, as in Eq.~(\ref{eqn:q2}).
\end{figure}

\vspace{1.5mm}
\hspace{-3.5mm}\textbf{Evolution over a quarter period}. At $t=\pi/2\omega$, $\exp[-\imath \omega tn]$ becomes $(-\imath)^n$ and this can be compensated by taking the Fourier transform of the eigenfunctions, as shown in Appendix B. Consequently every wave function changes essentially to its corresponding momentum wave function over $\frac{1}{4}T$. This result emerges directly from the propagator and is discussed in the next section.

\section{The propagator}
\label{sec:prop}
In principle, the problem of the evolution of wave functions is solved by finding the propagator $K(x,x',t)$ such that, for any initial wave function $\psi(x)$, its evolution at time $t$ is
\begin{equation} \label{eqn:propGen}
\psi(x,t)=\int_{-\infty}^{\infty}K(x,x',t) \psi(x') dx'.
\end{equation}
Derivations of the propagator for the oscillator have been given in many texts [for example, Ref.~\onlinecite{M1}, Section 15.2] and papers\cite{Th} and a derivation is given in Appendix A. The result is 
\begin{eqnarray}\label{eqn:prop}
K(x,x',t)&&=\frac{1}{\alpha\sqrt{2\pi\imath\sin \omega t}}\times \\
&&\exp[\frac{\imath}{2\alpha^2 \sin \omega t}((x^2 +x'^2)\cos \omega t - 2x x')],\nonumber
\end{eqnarray}
where $\alpha$ is our length scale $\sqrt{\hbar/m\omega}$. The propagator contains $t$ only as $\sin \omega t$ or $\cos \omega t$, and is therefore periodic with period $T=2\pi/\omega$, but there is an ambiguity in phase\cite{Th,L} due not only to the ambiguity in the sign of $\sqrt{\sin \omega t}$ but also to the singularity in the exponent whenever $\sin \omega t =0$. This ambiguity is resolved in Appendix A; the result is that an extra factor of $-\imath$ must be included each time $\sin \omega t$ goes through zero.

A simple result that can readily be derived from the propagator is that every Gaussian wave function remains Gaussian as it evolves.

The propagator also gives a simple result after any quarter period of evolution: at time $t=T/4$,
\begin{equation} \label{eqn:quartPeriod}
\psi(x,T/4)=\frac{1}{\alpha\sqrt{2\pi\imath}}\int_{-\infty}^{\infty} \exp[-\imath\frac{ x x'}{\alpha^2}]\psi(x') dx'.
\end{equation}
Because of the oscillator's symmetry between position and momentum, it helps to use the dimensionless variables $\xi=x/\alpha$ and $\rho=\alpha p/\hbar$ when taking Fourier transforms of wave functions. Write $\psi(x)=\Psi(x/\alpha)$ and define its Fourier transform as
\begin{equation}\label{eqn:FT}
\Phi(\rho)=\frac{1}{\sqrt{2\pi}}\int_{-\infty}^{\infty}\exp(-\imath \rho \xi) \Psi(\xi) d\xi.
\end{equation}
Then $\psi(x,T/4)=\exp(-\imath\pi/4)\,\Phi(x/\alpha)$. Thus, any state evolves over a quarter period into a state whose spatial wave function equals the Fourier transform of the initial wave function (apart from a phase change of $-\pi/4$).

Thus the propagator can be used to draw out some general features of the evolution. For application to specific initial wave functions, however, the integral in Eq.~(\ref{eqn:propGen}) is difficult to calculate numerically because the integrand has rapid oscillations in some regions. Other methods are used for numerical work (such as using an expansion of the wave function over the energy eigenfunctions). Furthermore, the integral can be evaluated analytically only for very few forms of initial wave function. Other than for an initial Gaussian (or higher energy eigenfunctions) those that can be evaluated give results in terms of special functions from which it is difficult to deduce the general features. For example, an initial square wave function can be evaluated in terms of a combination of error-functions with complex arguments.\cite{Anal} Note that the square wave function is too singular to be subject to most of the analysis below because its energy is infinite.

\section{Dynamics of the oscillator}
\label{sec:dynamics}
We will use the notation\cite{A} 
\begin{equation}
\label{eqn:DA}
D_t \hat{A}=\frac{\partial \hat{A}}{\partial t}+\frac{\imath}{\hbar}[\hat{H},\hat{A}].
\end{equation}
Then $D_t \hat{A}$ is the `total time-derivative' of the operator $\hat{A}$, which adds to the partial derivative a term that takes account of the evolution of the state under the Hamiltonian $\hat H$ in such a way that
\begin{equation}
\label{eqn:expDA}
d_t\langle\hat{A}\rangle=\langle D_t \hat{A}\rangle
\end{equation}
for any state and any operator $\hat{A}$, where $d_t$ stands for the usual time-derivative $d/dt$. It is easy to see that
\begin{equation}
\label{eqn:DAB}
D_t (\hat{A}\hat{B})=(D_t \hat{A})\hat{B}+\hat{A}(D_t \hat{B}),
\end{equation}
similar to differentiation of a function.

For the oscillator Hamiltonian in Eq.~(\ref{eqn:ham}), it is simple to calculate that
\begin{equation}\label{eqn:motion}
D_t \hat{x}=\hat{p}/m,  \hspace{5mm} D_t \hat{p}=-m\,\omega^2 \hat{x},
\end{equation}
similar to the equations of motion for the classical oscillator. These equations will be used to calculate how various expectation values evolve for the oscillator.

\vspace{1.5mm}
\hspace{-3.5mm}\textbf{Motion of the centroid}. 
We will refer to the expectation value of position as the {\it centroid} of the wave function. From Eqs.~(\ref{eqn:expDA}) and (\ref{eqn:motion}) 
\begin{equation}\label{eqn:motion1}
d_t \langle\hat{x}\rangle=\langle\hat{p}\rangle/m,  \hspace{5mm} d_t \langle\hat{p}\rangle=-m\,\omega^2 \langle\hat{x}\rangle.
\end{equation}
These are the usual classical equations for the oscillator with the solution $\langle\hat{x}\rangle=A \cos \omega (t-t_0)$ and $\langle\hat{p}\rangle=-m \omega A \sin \omega (t-t_0)$, where $t_0$ is the time of maximum displacement $A$. This is Ehrenfest's result that the centroid follows classical sinusoidal motion.

\section{Evolution is independent of the centroid}
\label{sec:indy}
Here we will show that the evolution of any wave function is independent of the motion of the centroid, except for a simple effect on the phase of the wave function. This greatly simplifies the problem of finding the evolution of an arbitrary wave function: one need only consider the evolution of wave functions whose centroid sits at the origin and does not move. Given any initial wave function $\psi(x)$, we can find the initial position $x_0=\langle \hat x \rangle_\psi$ and momentum $p_0=\langle \hat p \rangle_\psi$ of its centroid. Then the complete trajectory of the centroid follows from the classical equations of motion: $d_t x=p/m$ and $d_t p=-m\omega^2 x$. The centroid position is $\bar x(t)=x_0 \cos \omega t +(p_0/m\omega)\sin \omega t$ and momentum is $\bar p(t)=p_0 \cos \omega t -m\omega x_0\sin \omega t$. We can produce from $\psi(x)$ a wave function $\phi(x)=\exp(-\imath p_0 x/\hbar)\psi(x+x_0)$ whose centroid sits at the origin with no momentum. If $\phi(x,t)$ is the evolution of this initial wave function, and
\begin{equation}\label{eqn:psiphi}
\psi(x,t)=\exp[\frac{\imath }{\hbar}\bar p (t) (x-\frac{\bar x}{2})]\phi(x-\bar x (t),t),
\end{equation}
then, with $\xi=x-\bar x (t)$, it is not difficult to show that  
\begin{eqnarray}\label{eqn:SchOp}
&&[-\frac{\hbar^2}{2m}\frac{\partial^2 }{\partial x^2}+\frac{1}{2}m\omega^2 x^2 -\imath \hbar \frac{\partial}{\partial t}] \psi(x,t)\\ \nonumber
=&&\exp[\frac{\imath }{\hbar}\bar p (t) (x-\frac{\bar x}{2})][-\frac{\hbar^2}{2m}\frac{\partial^2 }{\partial \xi^2}+\frac{1}{2}m\omega^2 \xi^2 -\imath \hbar \frac{\partial}{\partial t}] \phi(\xi,t).
\end{eqnarray} 
That is, if $\phi(x,t)$ satisfies Schr\"odinger's equation, then so does $\psi(x,t)$.

\vspace{1.5mm}
\hspace{-3.5mm}\textbf{Displaced eigenstates}. 
As a simple example, consider the displaced ground state $\psi(x)=\exp[-(x-a)^2/2\alpha^2]$, where $a$ is any real constant. The centroid will follow $\bar x(t)=a \cos \omega t$, $\bar p(t)=-m\omega a\sin \omega t$. With $\xi=x+a$ we have $\phi(\xi,0)=\exp[-\xi^2/2\alpha^2]$, which is the ground state whose evolution is $\phi(\xi,t)=\exp[-\xi^2/2\alpha^2-\imath\omega t/2]$. Therefore the evolution of $\psi$ is, from Eq.~(\ref{eqn:psiphi}),
\begin{equation}\label{eqn:dgs}
\psi(x,t)=\exp[\imath\theta-(x-a\cos \omega t)^2/2\alpha^2],
\end{equation}
where $\theta=-a\sin\omega t\,\,(x-\frac{1}{2}a\cos \omega t)/\alpha^{2}-\frac{1}{2}\omega t$.

The same is true of the excited states. As shown in every textbook, the $n^{\mathrm{th}}$ eigenstate has the form $\psi_n(x,t)=\exp(-x^2/2\alpha^2)H_n(x/\alpha)\exp[-\imath(n+\frac{1}{2})\omega t]$ where $H_n(z)$ is the $n^{\mathrm{th}}$ Hermite polynomial. It then follows that there is this much larger class of displaced excited states, each defined by $n$ and the initial position $\bar x_0$ and momentum $\bar p_0$ of the centroid, that have the evolution
\begin{equation}
\Psi_n(x,t)=\exp[\imath\,\theta_n-\frac{1}{2}\xi^2/\alpha^2]H_n(\xi/\alpha),
\end{equation}
where $\xi=x-\bar x(t)$ and $\theta_n=[\bar p (t) \xi+\frac{1}{2}\bar p (t)\bar x (t)-E_n t]/\hbar$. Here $E_n=\hbar\omega(n+\frac{1}{2})$ and $\bar x(t)$, $\bar p(t)$ give the classical evolution of the centroid, $\bar x(t)=x_0 \cos \omega t +(p_0/m\omega)\sin \omega t$ and $\bar p(t)=p_0 \cos \omega t -m\omega x_0\sin \omega t$. For each of these states the magnitude $|\Psi_n(x-\bar x,t)|$ does not change with time, while the centroid $\bar x$ follows a classical oscillation.

\section{Evolution of the second moments}
\label{sec:2ndMoments}
It is worthwhile to study the second moments of the wave function,
\begin{eqnarray}\label{eqn:deltas}
\Delta_x^2 &=&\langle\hat x ^2\rangle -\langle\hat x\rangle^2,\\
\Delta_p^2 &=&\langle\hat p ^2\rangle -\langle\hat p\rangle^2,\\
\Delta_{xp} &=&\small{\frac{1}{2}}\langle\hat p\hat x+\hat x\hat p\rangle -\langle\hat x\rangle\langle\hat p\rangle,
\end{eqnarray}
because $\Delta_{x}$ and $\Delta_{p}$ are measures of the spread in position and momentum of the state, while $\Delta_{xp}$ is a measure of the correlation between position and momentum, but also because these moments enable insight into the energy and uncertainty properties of the state. We can use the operator equations (\ref{eqn:DAB}) and (\ref{eqn:motion}), to derive
\begin{eqnarray}\label{eqn:OpMotion2}
D_t \hat x ^2 &=& (\hat p\hat x+\hat x\hat p) /m ,\\
D_t \hat p ^2 &=& -m\omega^2(\hat p\hat x+\hat x\hat p) ,\\
D_t (\hat p\hat x+\hat x\hat p) &=& 2(\hat p ^2 /m - m\omega^2\hat x ^2) ,
\end{eqnarray}
and there is a similar set of equations for $\langle\hat x ^2\rangle , \langle\hat p ^2\rangle$ and $\langle\hat p\hat x+\hat x\hat p\rangle$, as well as for $\langle\hat x\rangle^2 , \langle\hat p\rangle^2$ and $2\langle\hat p\rangle\langle\hat x\rangle$. Hence
\begin{eqnarray}\label{eqn:duv}
d_t \Delta_x^2 &=& 2\Delta_{xp}/m,\\
d_t \Delta_p^2&=&-2m\omega^2\Delta_{xp},\label{eqn:duv2}\\
d_t \Delta_{xp}&=&\Delta_p^2/m-m\omega^2\Delta_x^2.\label{eqn:duv3}
\end{eqnarray}
It follows that $d_t^2 \Delta_{xp} = -4\omega^2 \Delta_{xp}$
and therefore $\Delta_{xp}$ varies sinusoidally with twice the frequency of the oscillator. We can take
\begin{equation}\label{eqn:v}
\Delta_{xp}=\hbar A \sin 2\omega (t-t_0),
\end{equation}
where $t_0$ is a time when $\Delta_{xp}$ is zero. Then, integrating Eqns.(\ref{eqn:duv}) and (\ref{eqn:duv2}),
\begin{eqnarray}\label{eqn:us}
\Delta_x^2 &=& \alpha^2[\epsilon-A \cos 2\omega(t-t_0)] \\
\Delta_p^2 &=& (\hbar/\alpha)^2[\epsilon+A \cos 2\omega(t-t_0)]. \nonumber
\end{eqnarray}
The constant $\epsilon$ must be the same in each of these equations to satisfy Eq.~(\ref{eqn:duv3}). These basic properties are illustrated in Fig.~2 showing $\Delta_x^2/\alpha^2$, $\alpha^2\Delta_p^2/\hbar^2$ and $\Delta_{xp}/\hbar$ over two periods of the oscillator, for the sum of two Gaussians shown in Fig.~1. More detail is given in Section \ref{sec:Examples}.

\begin{figure}[h!]
\centering
\includegraphics{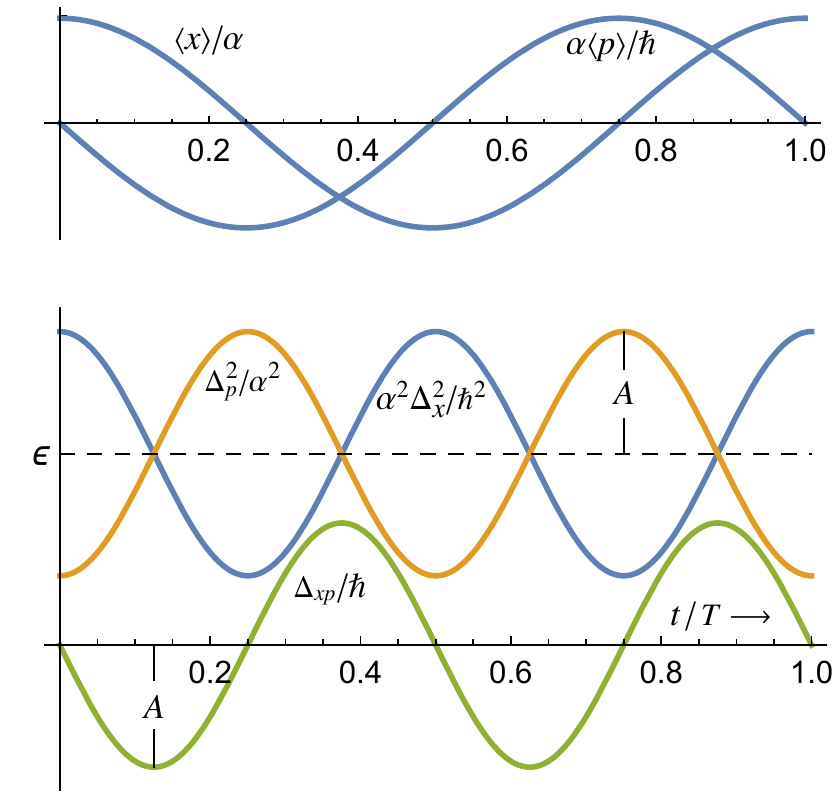}
\caption{\label{uv}}
The oscillations of the first and second moments of the system described in Fig.~1. The top graph shows the position $\langle \hat x\rangle$ and momentum $\langle \hat p\rangle$ of the centroid, each scaled to be dimensionless, plotted against the time in units of the period $T=2\pi/\omega$ of the oscillator. The lower graph shows $\Delta_x^2$, $\Delta_p^2$ and $\Delta_{xp}$, each scaled to be dimensionless. The amplitude of the oscillations in each of these three curves is $A$. The vertical scale for the centroid (top graph) bears no relation to that for the second moments, although the time is aligned. In general, the phase of the oscillation of the centroid need not align with that of the second moments.\end{figure}

The question of how to calculate the constants $\epsilon$, $A$ and $t_0$ in terms of $\Delta_x^2$, $\Delta_p^2$ and $\Delta_{xp}$ will be taken up at the end of the next section.

In summary, for any wave function, the evolution of the second moments can be simply calculated from their values at any initial time. The spreads $\Delta_x ^2$ and $\Delta_p ^2$ oscillate sinusoidally about fixed values at twice the frequency of the oscillator, always differing in phase by $180\deg$. The correlation $\Delta_{xp}$ oscillates about zero with phase differing by $90\deg$ from the others. Although $\Delta_x$ and $\Delta_p$ can be interpreted as spreads about the centroids, their evolution does not depend on the motion of the centroid.

\section{Energy and uncertainty}
\label{uncertain}
We define the {\it quantal energy} to be $E_q=\langle \hat H \rangle-E_c$ where $E_c=\langle \hat p \rangle^2/2m + m\omega^2\langle \hat x \rangle^2/2$ is the classical energy of the centroid. Thus
\begin{equation}\label{eqn:Eq}
E_q = \frac{1}{2m}\Delta_p^2 + \frac{1}{2}m\omega^2\Delta_x^2=\hbar\omega\epsilon.
\end{equation}
The constant $\epsilon=E_q/\hbar\omega$ is then the quantal energy in units of $\hbar\omega$. Both $E_q$ and $E_c$ remain constant.

Although the initial values of all three second-order moments are needed to calculate the evolution of any of them, the combination
\begin{equation}\label{eqn:K}
K^2 = (\Delta_x^2 \Delta_p^2 - \Delta_{xp}^2)/\hbar^2
\end{equation}
remains constant, as is easily verified using Eqs.~(\ref{eqn:duv}-\ref{eqn:duv3}). Also, $K$ is subject to the inequality $K\geq \frac{1}{2}$, sometimes known as the Schr\" odinger-Robinson inequality [Ref.\onlinecite{M1}, Eq.~(10.57)]. It is stronger than (and includes) Heisenberg's uncertainty relation $\Delta_x \Delta_p \geq \hbar/2$. An alternative proof of the the inequality will emerge in Section \ref{sec:scale}.

From Eqs.~(\ref{eqn:v}) and (\ref{eqn:us}) follows the relation
\begin{equation}\label{eqn:eAK}
\epsilon^2=A^2+K^2,
\end{equation}
and therefore $\epsilon\geq K$. Also $\epsilon=K$ if and only if $A=0$ which implies no oscillations in the second moments.

From Eq.~(\ref{eqn:K}), $\Delta_x^2 \Delta_p^2 =\hbar^2 K^2 + \Delta_{xp}^2$ and $\Delta_{xp}$ oscillates between $-\hbar A$ and $\hbar A$. Therefore the `uncertainty product' $\Delta_x \Delta_p$ oscillates between $\hbar K$ and $\hbar\epsilon$. Thus
\begin{equation}
\epsilon\;\geq \;\Delta_x \Delta_p/\hbar\;\geq K\;\geq\; \small{\frac{1}{2}}.
\end{equation}
A physical manifestation of $K$ is that it is the minimum value that $\Delta_x \Delta_p/\hbar$ takes as it oscillates.

Returning to the question of how to calculate the constants $\epsilon$, $A$ and $t_0$ in terms of the initial moments, $\epsilon$ and $K$ can be found from Eqs.~(\ref{eqn:Eq}) and (\ref{eqn:K}). Then $A$ follows from Eq.~(\ref{eqn:eAK}) while Eqs.~(\ref{eqn:v}) and (\ref{eqn:us}) give (taking $t=0$)
\begin{eqnarray}
A\sin 2\omega t_0 &=& -\Delta_{xp}/\hbar \\
A\cos 2\omega t_0 &=& \frac{1}{2}(\frac{\alpha^2}{\hbar^2}\Delta_p^2-\frac{1}{\alpha^2}\Delta_x^2).
\end{eqnarray}

\section{Changing the scale of a state}
\label{sec:scale}
We saw that the evolution of any state can be reduced to that of a state that remains at rest at the origin. This transformation involved only a spatial translation and a change in phase that depends only on the first moments of the state. Now we show that the evolution of these states can be further reduced to states whose second moments are constant, which we refer to as \textit{stable} states. 

\vspace{2mm}
\hspace{-3.5mm}\textbf{Stable states}. From Eqs.~(\ref{eqn:v}-\ref{eqn:us}), having constant second moments requires $A=0$. This implies that: $K=\epsilon$, $\Delta_{xp}=0$, $\Delta_x^2=\alpha^2 K$, $\Delta_p^2=\hbar^2 K/\alpha^2$, $\Delta_x \Delta_p=\hbar K$ and these values will be maintained as the state evolves. Any state with $\Delta_{xp}=0$ (which includes any with a real wave function) is stable if and only if $\Delta_x/\alpha=\alpha\Delta_p/\hbar$. Although the second moments of stable states do not change with time, the wave functions do change to some extent. Energy eigenfunctions are stable, but their phase does change with time. If $K=\frac{1}{2}$ the stable state has $\epsilon=\frac{1}{2}$ and therefore must be the ground state Gaussian.

\vspace{2mm}
\hspace{-3.5mm}\textbf{Changing scale}. In the case of a free particle, changing scale is a simple matter: if $\psi(x,t)$ satisfies Schr\"odinger's equation $\imath\hbar\,\partial_t\psi=-(\hbar^2/2m)\partial^2_x \psi$ then so does $\psi(sx,s^2t)$. Thus if we change the spatial scale, we still have a solution (with a different time scale). For the oscillator, we can find a solution with a different spatial scale, but this generally requires an added phase (that varies with $x$ and $t$) and a change the time scale, now to a nonlinear function of the original time.

\vspace{2mm}
\hspace{-3.5mm}\textbf{Any state can be transformed to a stable form}. For a state $\psi(x)$ with $\Delta_{xp}=0$, which includes any with a real wave function (as shown in Appendix D), we can find the scale $s$ such that $\psi(sx)$ has lower energy than any other scale, and this lowest-energy state is stable. An example is worked out for the triangular wave function in Example 2 of Section \ref{sec:Examples}. Otherwise, when $\psi(x)$ has $\Delta_{xp}\neq 0$ we can find a phase change that reduces $\Delta_{xp}$ to zero and then change the scale. The details are in Appendix C and the result is that
\begin{equation}\label{eqn:psi2phi}
\phi(x)=\exp(-^1_{\bar 2}\imath x^2\!/b^2)\,\psi(sx),
\end{equation}
is stable if $s=\Delta_x/\alpha\sqrt{K}$ and $b^2=\alpha^2 \hbar K/\Delta_{xp}$, where $\Delta_x$, $\Delta_{xp}$ and $K$ refer to $\psi(x)$. In fact $K$ is unchanged by this transformation. 

Every stable state has $\epsilon=K$ and its energy cannot be less than $\small{\frac{1}{2}}\hbar\omega$. Therefore every state must have $K\geq \small{\frac{1}{2}}$, which proves the strong uncertainty inequality.

\vspace{2mm}
\hspace{-3.5mm}\textbf{Evolution of a state from its stable form}. The inverse of Eq.~(\ref{eqn:psi2phi}), expressing $\psi(x)$ in terms of $\phi(x)$, is $\psi(x)=\exp(\frac{1}{2}\imath x^2/b^2s^2)\phi(x/s)$, and this is the starting point to find the evolution $\psi(x,t)$ of $\psi(x)$ from the evolution $\phi(x,t)$ of $\phi(x)$. The time $t$ in $\phi(x,t)$ must be distorted by replacing $t$ by some function $\tau(t)$. Thus, if we insert
\begin{equation}\label{eqn:psiEvolve}
\psi(x,t)=\sqrt{\frac{\Delta_0}{\Delta_x}}\exp[\frac{\imath}{\hbar}\Delta_{xp}\frac{x^2}{2\Delta_x^2}]\,\phi\big(\frac{\sqrt{K}\alpha\,x}{\Delta_x},\tau(t)\big)
\end{equation}
into Schr\"odinger's equation (where the factor $\Delta_x^{-1/2}$ is clearly required to preserve normalization), we find that $\psi(x,t)$ satisfies Schr\"odinger's equation if $\phi(x,t)$ does, provided that $d\tau/dt=K\alpha^2/\Delta_x^2$ and $\Delta_x$, $\Delta_{xp}$ evolve as in Eqns.(\ref{eqn:duv}-\ref{eqn:duv3}). [Similarly we can find the evolution of $\exp(-\imath x^2/2a^2)\psi(x)$ if we know that of any $\psi(x)$.]

\vspace{2mm}
\hspace{-3.5mm}\textbf{The distorted time $\tau(t)$} is found by integrating $d\tau/dt=K\alpha^2/\Delta_x^2$, inserting $\Delta_x^2$ from Eq.~(\ref{eqn:us}), to give
\begin{equation}\label{eqn:tau}
\tau(t)=\frac{1}{\omega}\arctan[\frac{\epsilon+A}{K}\tan\omega (t-t_0)].
\end{equation}
When $A=0$ then $\epsilon=K$ and hence $\tau(t)=t-t_0$. An example of $\tau(t)$ is shown in Fig.~3. As required by the periodicity properties, $\tau$ equals $t-t_0$ at multiples of $T/4$. [As an alternative, Eq.~(\ref{eqn:tau}) is equivalent to $\exp 2\imath\omega\tau=(1+\imath z)/(1-\imath z)$, where $z=[(\epsilon+A)/K] \tan \omega(t-t_0)$.]

\begin{figure}[h!]
\centering
\includegraphics{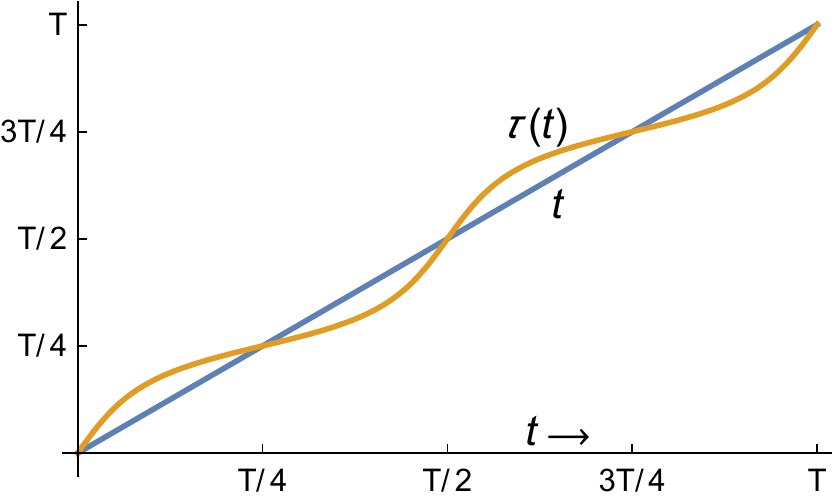}
\caption{\label{tau}}
The distorted time $\tau(t)$ compared with the actual time $t$ elapsed since a time when $\Delta_{xp}=0$. In the case shown, $K=1$, $A=1$ and therefore $\epsilon = \sqrt{2}$.
\end{figure}

\hspace{-3.5mm}\textbf{The evolution of a scaled stable state} $\phi(\bar s x,0)$ from $\phi(x,t)$ is given by Eq.~(\ref{eqn:psiEvolve}) initially taking $\Delta_{xp}=0$. Then $\Delta_{xp}=\hbar A \sin 2\omega t$ and $\Delta_x^2 = \alpha^2(\epsilon-A \cos 2\omega t)$, so that $\psi(x,0)=\phi(\bar s x,0)$, where $\bar s^{\,2}=K/(\epsilon-A)=(\epsilon+A)/K$. 

\vspace{2mm}
\hspace{-3.5mm}\textbf{Squeezed states} result from changing the scale of the ground state. They have $K=\frac{1}{2}$ and can be defined by the amplitude $A$ of the moments. The evolution of the ground state is $\phi(x,t)=\exp(-\frac{1}{2}x^2/\alpha^2-\frac{1}{2}\imath\,\omega t)$ and so, from Eq.~(\ref{eqn:psiEvolve}), the evolution of the squeezed state is
\begin{equation}\label{eqn:sq}
\psi(x,t)=\sqrt{\frac{\Delta_0}{\Delta_x}}\exp[\big(\frac{\imath}{\hbar}\Delta_{xp}- \frac{1}{2}\big)\frac{x^2}{2\Delta_x^2}-\small{\frac{1}{2}}\imath\,\omega\tau(t)],
\end{equation}
where $\Delta_x=\alpha(\epsilon-A\cos2\omega t)^{1/2}$ with $\Delta_0=(\Delta_x)_{t=0}$, $\Delta_{xp}=\hbar A\sin 2\omega t$, $\omega\tau(t)=\arctan[2(\epsilon +A)\tan\omega t]$ and $\epsilon=(A^2 +\frac{1}{4})^{1/2}$. This is easily extended to the evolution of excited energy eigenstates after a change of scale. 

For a study of the evolution of squeezed states using the propagator, see Ref.~\onlinecite{Gersch}. Squeezed states are important in quantum optics, where the mathematics of the quantized oscillations of the electromagnetic field mirror that of the harmonic oscillator.\cite{GCS}

\vspace{2mm}
\hspace{-3.5mm}\textbf{Why change scale?} Before carrying out a numerical calculation of the evolution of a wave function $\psi(x)$ (such as the triangular wave function in Example 2) it is useful to first transform the state to a stable form. Then the results can easily be applied to all the wave functions of the form $\psi(sx)$; all that is required is a relabelling of $x$ and $t$ and an easily calculated phase factor.

More generally, calculations of physical processes often use a basis of states. The scale of the system may change significantly with time, and then it may give considerable advantage to allow the basis to change in scale to match the changing scale of the system.

\section{Examples of evolution}
\label{sec:Examples}

\hspace{-3.5mm}\textbf{Example 1} is a superposition of two ground-state Gaussian packets displaced from the origin, as shown in Figs.~1 and 2. We will discuss how to calculate the evolution of the wave function and its first and second moments. The Gaussians start at rest, one centered on $x=a_1$ while the other, with amplitude $b$ relative to the first, centered on $x=a_2$.  Thus the initial wave function is $\psi(x)=\psi_1(x)+b\,\psi_2(x)$, where $\psi_1(x)=\phi_0(x-a_1)$ and $\psi_2(x)=\phi_0(x-a_2)$, where $\phi_0(x)=\exp(-x^2/2\alpha^2)$. The trajectory of the centroids will be $x_k(t)=a_k \cos(\omega t)$ and $p_k(t)=a_k \cos(\omega t)$ for $k=1,2$.
From Eq.~(\ref{eqn:psiphi}), $\psi_k(x)$ will evolve to
\begin{equation}\label{eqn:psi1}
\psi_k(x,t)=\exp[\imath\,\theta_k(x,t)-\frac{1}{2\alpha}(x-x_k(t))^2],
\end{equation}
where $\theta_k(x,t)=p_k(t)(x-\frac{1}{2}x_k(t))/\hbar-\frac{1}{2}\omega t$. Then $\psi(x,t)=\psi_1(x,t)+b\psi_2(x,t)$ is the evolution of the superposition $\psi(x)$ and is depicted in Fig.~1. The integrals for the initial second moments can be carried out exactly and the results contain only powers and exponentials (but are best done with a computer algebra/calculus program). Once the initial moments are calculated, Eqs.~(\ref{eqn:v}) and (\ref{eqn:us}) give the evolution of the moments as shown in Fig.~2.

\vspace{2mm} 
\hspace{-3.5mm}\textbf{Example 2}. Another example is the triangular wave function, with $\psi(x)=1-|x|/a$ for $|x|\leq a$. In contrast to Example 1, the moments are easily calculated and reveal a lot about the evolution, but the exact evolution is not easily available. The wave function is real and it follows (see Appendix D) that initially \mbox{$\langle\hat x\rangle\!=\!\langle\hat p\rangle\!=\!\langle(\hat x\hat p\!+\!\hat p\hat x)\rangle\!=\!0$}. It is easy to calculate that $\langle\hat x^2\rangle\!=\!a^2/10$ and $\langle\hat p^2\rangle\!=\!3\hbar^2/a^2$. Thus we have the initial values $\Delta_x^2=a^2/10$, $\Delta_p^2=3\hbar^2/a^2$ and $\Delta_{xp}=0$, from which we can calculate the constants $K^2\!=\!3/10$, $\epsilon\!=\!\frac{1}{2}(\lambda^2/10\!+\!3/\lambda^2)$ and $A\!=\!\frac{1}{2}(\lambda^2/10\!-\!3/\lambda^2)$, where $\lambda=a/\alpha$ is the only non-trivial parameter for this wave function. Then the minimum energy, with $\epsilon=K=(3/10)^{1/2}\approx 0.5477$ occurs when $A=0$ and requires that $\lambda=30^{1/4}\approx 2.34$. This is the stable triangular wave function. It is the wave function most similar (in this sense) to the ground state of the oscillator, and its second moments will remain constant. These two wave functions are compared in Fig.~4.

\begin{figure}[h!]
\centering
\includegraphics{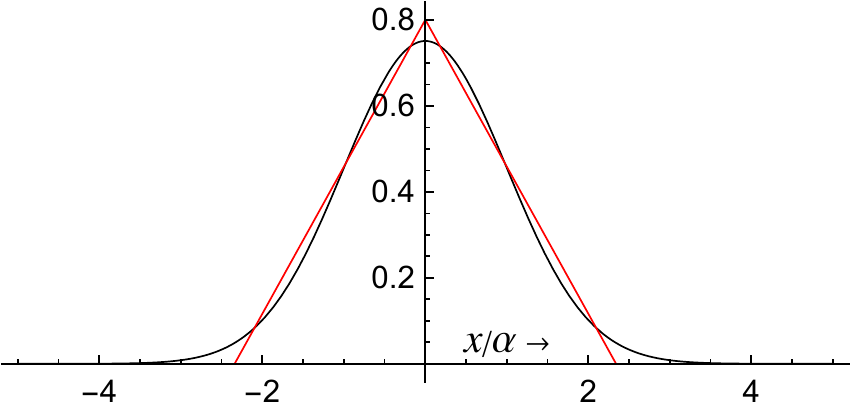}
\caption{\label{TriGauss}}
The stable (i.e. minimum-energy) triangular wave function ($a\approx 2.34\alpha$) compared with the ground-state wave function. Both wave functions are normalized to $\int_{-\infty}^{\infty}\psi^2\,dx=1$.
\end{figure}

The Fourier transform of the triangular wave function can also be calculated exactly and gives
\begin{equation}
\psi(x,T/4)=e^{\imath\pi/4}\frac{4\sin^2(\frac{1}{2}ax)}{\sqrt{2\pi}\,ax^2}.
\end{equation}
For the minimum-energy triangular wave function, the absolute value of this Fourier transform does not differ greatly from the original triangle (as can be seen from the top graph in Fig.~5) and since $\Delta_x$ is constant we might expect that the whole evolution remains close to the triangle. To be more specific, we can express the normalized initial state $\psi$ as a linear combination of the ground state $\psi_0$ and a part $\chi$ orthogonal to $\psi_0$: $\psi=c_0 \psi_0+c_\chi \chi$, with $\langle\psi |\psi\rangle=1$, $\langle\psi_0 |\psi_0\rangle=1$, $\langle\chi |\chi\rangle=1$ and $\langle\psi_0 |\chi\rangle=0$. Finding $c_0$ and $c_\chi$ requires only a numerical integration of $\psi \psi_0$ and the result is that $c_0^2\approx0.9953$ and therefore $c_\chi^2\approx0.0047$. Thus the non-Gaussian part of $\psi$ contributes less than $1/200$ to the probability distribution and this will be maintained throughout the evolution. The Gaussian part remains unchanged (apart from its phase) and all the deviations from the Gaussian come from this small $\chi$. 

This is borne out by a calculation of the evolution, first expanding the triangle as a sum of oscillator eigenfunctions (by numerical integration) and then including the time-dependence of the eigenfunctions. Figure 5 shows the result for $|\psi(x,t)|$ over the first quarter period. 

\begin{figure}[h!]
\centering
\includegraphics{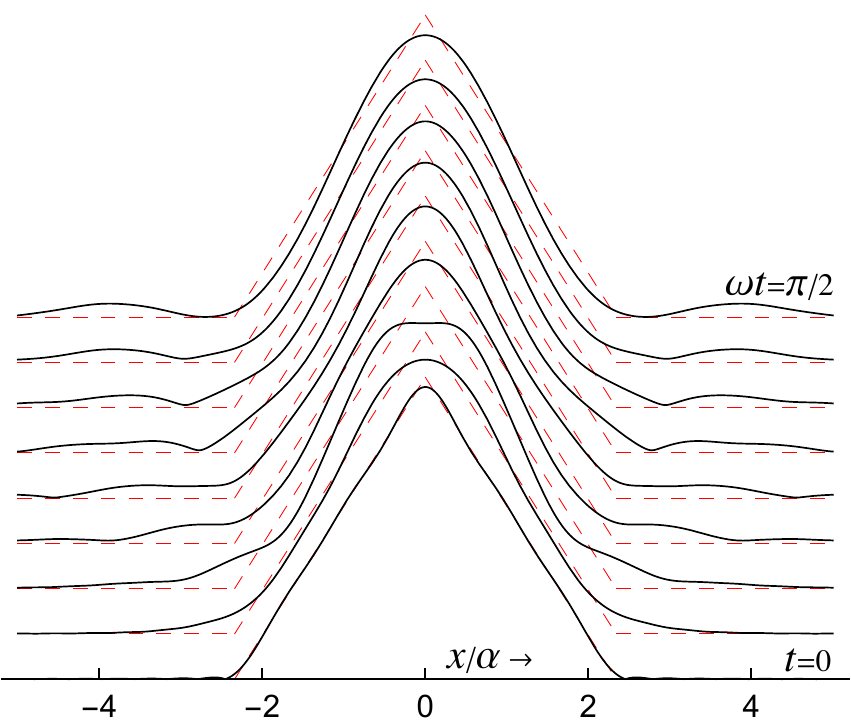}
\caption{\label{TriEv}}
The evolution of the minimum-energy triangular wave function over a quarter period. The time interval between each plot is $T/32$ and $|\psi|$ is shown (solid curve) while the original triangular wave function is shown (dashed) at each time for comparison.
\end{figure}

For a wider initial triangle, that is $a>30^{1/4}\alpha$, the result is similar except that $\Delta_x$ decreases over the first quarter period. The evolving wave function stays close to an evolving Gaussian, which is now a squeezed state rather than a coherent state.

\section{Conclusions}
\label{sec:Conclusion}
In principle, the evolution of oscillator wave functions is provided by the known propagator, but only a small set of initial wave functions allow the integral over the propagator to give a simple result and even then the main attributes of the evolution may be hidden in the mathematical formulae. It is useful, therefore, to investigate some general features of the evolution.

One simple feature is the periodicity. After one quarter of the classical period $T=2\pi/\omega$, any wave function evolves to one that can be simply expressed in terms of the Fourier transform of the original wave function. A simpler result applies over half a period: $\psi(x,\frac{1}{2}T)=-\imath\psi(-x,0)$. Applying this result twice gives $\psi(x,T)=-\psi(x,0)$. Thus the true period of the quantum oscillator is twice that of the classical oscillator; evolution over $2T$ returns the wave function exactly to its original. These results imply that, for any initial wave function, the evolution over any time interval can be found from that through the first half period. If the initial wave function is real, only the first quarter period is needed.

The simplest fact about the continuous evolution is Ehrenfest's result that the centroid strictly follows classical evolution; that is, sinusoidal oscillation that can be easily calculated from the initial values. The evolution of the whole wave function relative to the centroid, is essentially independent of that of the centroid, and therefore we need consider only states that remain centered on the origin (the center of force of the oscillator); that is, states with $\langle\hat x\rangle=\langle\hat p\rangle=0$. Any other wave function can be evolved by first applying a simple transformation that depends only on $\langle\hat x\rangle$ and $\langle\hat p\rangle$. This greatly simplifies the practical problem of evolution of an oscillator.

Without calculating the full evolution, a great deal about the behaviour of wave functions can be obtained from the three second moments $\Delta_x^2$, $\Delta_p^2$ and $\Delta_{xp}$, which form a closed set: given their initial values we can find their evolution, independent of any other properties of the wave function. The energy can be separated into the energy $E_c$ of the centroid (the energy of the equivalent classical oscillator) which depends only on the first moments, and the quantal energy $E_q$, which is the remaining part of $\langle\hat H\rangle=E_c+E_q$. The quantal energy can be expressed in terms of second moments, $E_q=\Delta_p^2/2m+\frac{1}{2}m\omega^2\Delta_x^2$, and $E_q\geq \frac{1}{2}\hbar\omega$. Both $E_c$ and $E_q$ remain constant. The evolution is also constrained by the constancy of the combination $K^2=(\Delta_x^2\Delta_p^2-\Delta_{xp}^2)/\hbar^2$. This constant $K$ is subject to the inequality $K\geq \frac{1}{2}$, which contains (but is stronger than) the traditional Heisenberg uncertainty relation. The extreme case of $K=\frac{1}{2}$ implies that the wave function is Gaussian, and is known as a squeezed state. Apart from the motion of the centroid and the choice of the origin of the time, squeezed states are characterized by a single parameter that defines the extent of the oscillations of the second moments. The case with no such oscillations is the \textit{coherent} state, equivalent to the ground state of the oscillator.

For some states, here called stable states, the second moments do not change with time. If we have the evolution of a stable state $\psi(x)$, then the evolution of any state $\psi(s x)$  can easily be written down. Also, from any state $\psi(x)$ one can find a stable state through a transformation that involves only its second moments, and the evolution of $\psi(x)$ can be easily found from that of its stable state. Thus the evolution of every oscillator wave function (for which the second moments exist) is simply related to that of a stable state at rest at the origin.

The main focus here has been the basic principles; but the potential applications are spread widely in physics, because harmonic oscillations are ubiquitous. In practice, there will be other physical processes acting (such as radiation and environmental effects). These can be dealt with using a basis of energy eigenfunctions of the oscillator, but in some cases it is more effective to shift and scale the basis to match the changes in the object being studied. The evolution of such displaced and squeezed eigenstates was discussed in Sections (\ref{sec:indy}) and (\ref{sec:scale}).

In two or three dimensions all the results above can be applied in each dimension since the oscillator Hamiltonian can be separated as a sum of commuting terms. Thus, for example, in the case of an isotropic oscillator in three dimensions, the centroid of every state will travel along an elliptic orbit in a plane and the evolution of the wave function can be reduced to that of a state that remains centered on the origin.

\appendix
\section{The propagator from an invariant operator}

The propagator $K(x,x',t)$, as defined by Eq.~(\ref{eqn:propGen}), is the solution of Schr\"odinger's equation such that $K(x,x',t)\to \delta(x-x')$ as $t\to 0$ and has the symmetry property $K(x',x,-t)=K^*(x,x',t)$.\cite{M1}

If we can find an operator $\hat u$ such that $\hat u\to\hat x$ as $t\to 0$, then the solution of $\hat u\,\psi=x'\psi$, will have $\psi\propto \delta(x-x')$ as $t\to 0$, because $\delta(x-x')$ is the eigenfunction of $\hat x$ with eigenvalue $x'$. If $\hat u$ is also an \textit{invariant} operator, that is $D_t\hat u=0$ with $D_t$ defined in Eq.~(\ref{eqn:DA}), then\cite{A} $\psi$ will satisfy Schr\"odinger's equation $\hat H \psi=\imath\hbar\partial_t\psi$ apart from a possible time-dependent factor, and will give the $x$-dependence of the propagator.  

In the case of the oscillator, we can find such an operator $\hat u$ as a linear combination of $\hat x$ and $\hat p$. Thus if $\hat u=P(t)\hat x-X(t)\hat p$, then $D_t \hat u=(\dot P+m\omega^2 X)\hat x-(\dot X-P/m)\hat p$, so $\hat u$ will be invariant if $X(t),P(t)$ satisfy the classical equations of motion. Therefore, we take
\begin{equation}\label{eqn:propop}
\hat u=\cos\omega t \,\hat x-(m\omega)^{-1}\sin\omega t \,\,\hat p.
\end{equation}
Then the solution of $\hat u\psi=x'\psi$ is
\begin{equation}\label{propsi}
\psi(x,t)=f_1(x',t)\exp[\frac{\imath m\omega}{2\hbar\sin\omega t}(\cos\omega t\,x^2-2x'x)],
\end{equation}
and this gives the $x$-dependence of the propagator. The symmetry between $x$ and $x'$ then requires
\begin{equation}\label{propK}
K(x,x',t)=f_2(t)\exp[\frac{\imath m\omega}{2\hbar\sin\omega t}(\cos\omega t(x^2+x'^2)-2x'x)],
\end{equation}
and inserting this into Schr\"odinger's equation shows that $K(x,x',t)$ will be a solution provided that $d_tf_2(t)=-\omega\cot \omega t\,f_2(t)$ and therefore $f_2(t)=c/\sqrt{\sin \omega t}$. All that remains to be determined is the constant $c$ and this can be found by requiring that, for small $t$, $\int^\infty_{-\infty}K(x,x',t)\exp(-x^2/b^2)\,dx$ should approach unity as $b\to \infty$. This gives $f_2(t)=\sqrt{m\omega/2\pi\imath\hbar\sin\omega t}$. Making the phase precise, the propagator is therefore
\begin{equation} \label{eqn:propA}
K(x,x',t)=\frac{e^{-\imath\pi/4}\alpha^{-1}}{\sqrt{2\pi|\sin \omega t|}}\exp[\imath\frac{(x^2 +x'^2)\cos \omega t - 2x x'}{2\alpha^2 \sin \omega t}],
\end{equation}
agreeing with Eq.~(\ref{eqn:prop}). As discussed in Section \ref{sec:prop}, the phase given here is valid only for $0 <\omega t <\pi$ because of the singularity when $\sin\omega t=0$.

\vspace{2mm}
\hspace{-3.5mm}\textbf{Resolving the ambiguity in phase}. This analysis shows that a representation of the $\delta$-function is
\begin{equation}\label{eqn:delta}
\delta(x)=\lim_{\epsilon\to 0+}\frac{e^{-\imath\pi/4}}{\alpha\sqrt{2\pi\epsilon}}\exp\frac{\imath x^2}{2\epsilon\alpha^2},
\end{equation}
so that, for small $\omega t$, $K(x,x',t)\to \delta(x-x')$. For $\omega t=\pi-\epsilon$, Eq.~(\ref{eqn:propA}) becomes
\begin{eqnarray}
K(x,x',t) &\approx & \frac{e^{-\imath\pi/4}}{\alpha\sqrt{2\pi\epsilon}}\exp\frac{-\imath (x+x')^2}{2\epsilon\alpha^2}\\
&\to & e^{-\imath\pi/2}\delta(x+x')\;\;\textrm{as}\;\;\epsilon\to 0,
\end{eqnarray}
where we have used the complex conjugate of Eq.~(\ref{eqn:delta}) for the $\delta$-function. Hence, $\psi(x,t)\approx e^{-\imath\pi/2}\psi(-x,0)$, as required. Now if we pass through the singularity to time $\omega t=\pi+\epsilon$, Eq.~(\ref{eqn:propA}) gives $K(x,x',t)=\delta(x+x')$, which is incorrect as it involves a sudden jump in phase. To get the correct result, we must multiply $K(x,x',t)$ by $e^{-\imath\pi/2}=-\imath$. This extra phase is required each time $\omega t$ passes through a multiple of $\pi$. Therefore Eq.~(\ref{eqn:propA}) requires an extra factor of $(-\imath)^k$, where $k$ is the integer part of $\omega t/\pi$. This phase change can be better understood through Maslov theory and phase space geometry.\cite{L}

\section{The Fourier transform of eigenfunctions and evolution over $T/4$}

The standard lowering operator for the oscillator is $\hat a =(\hat x/\alpha+\imath \alpha \hat p/\hbar)/\sqrt{2}$ and the ground state $\psi_0$ has $\hat a \psi_0=0$ with the wave function $\psi_0(x)=\exp (-x^2/2\alpha^2)$. From this ground state, we can generate the infinite sequence of excited eigenstates by repeatedly applying the raising operator $\hat a^\dag$. Thus $\psi_{n+1}=c_n \hat a^\dag \psi_n$, where $c_n$ is some real number that could be taken to preserve normalisation, but is not relevant here.

We will now find a relation between $\psi_n(x)$ and its Fourier transform (or the momentum wave function). This relation arises from the symmetry between position and momentum in the oscillator, and it helps to use dimensionless variables $\xi=x/\alpha$ and $\rho=\alpha p/\hbar$ to make this symmetry more explicit. [For example, Schr\"odinger's equation becomes $\frac{1}{2}\omega(\hat\rho^2+\hat\xi^2)\psi=\imath\partial_t\psi$.] Write $\psi(x)=\Psi(x/\alpha)$ and define its Fourier transform as
\begin{equation}
\Phi(\rho)=\mathcal{F}\Psi=\frac{1}{\sqrt{2\pi}}\int_{-\infty}^{\infty}e^{-\imath \rho \xi} \Psi(\xi) d\xi.
\end{equation}
Then the ground state $\Psi_0(\xi)=\exp(-\frac{1}{2}\xi^2)$ has the Fourier transform $\Phi_0(\rho)=\exp(-\frac{1}{2}\rho^2)$. Now suppose we have some function $\Psi(\xi)$ and examine the Fourier transform of $\hat a^\dag \Psi(\xi)$.
We have $\hat a^\dag = (\xi-\partial_\xi)/\sqrt{2}$ and, integrating by parts,
\begin{eqnarray}
&&\int_{-\infty}^{\infty}e^{-\imath \rho \xi}(\xi-\partial_\xi) \Psi(\xi) d\xi 
= \int_{-\infty}^{\infty}e^{-\imath \rho \xi} (\xi-\imath\rho) \Psi(\xi) d\xi  \nonumber \\ 
&& \hspace{10mm} =-\imath (\rho-\partial_\rho) \int_{-\infty}^{\infty}e^{-\imath \rho \xi}\Psi(\xi) d\xi.
\end{eqnarray}
That is,
\begin{equation}\label{eqn:FF}
\mathcal{F}(\hat a^\dag\Psi)=-\imath\,\hat a^\dag(\mathcal{F}\Psi).
\end{equation}
We showed that $\mathcal{F}\Psi_0(\xi)= \Psi_0(\rho)$; then Eq.~(\ref{eqn:FF}) gives $\mathcal{F}\Psi_1(\xi)= -\imath\Psi_1(\rho)$. [If we take the first excited eigenfunction to be $\Psi_1(\xi)=\hat a^\dag \Psi_0(\xi)/\sqrt{2}=\xi\exp(-\frac{1}{2}\xi^2)$, then its Fourier transform is $\Phi_1(\rho)=-\imath\rho\exp(-\frac{1}{2}\rho^2)$.] This process can be continued, and an extra $-\imath$ is factored in at each step. Thus
\begin{equation}\label{eqn:F}
\mathcal{F}\Psi_n(\xi)=(-\imath)^n \Psi_n(\rho).
\end{equation}
More formally, if Eq.~(\ref{eqn:F}) holds for some $n$, then Eq.~(\ref{eqn:FF}) gives $\mathcal{F}(\hat a^\dag\Psi_n(\xi))=(-\imath)^{n+1} \hat a^\dag(\Psi_n(\rho))$ and therefore $\mathcal{F}\Psi_{n+1}(\xi)=(-\imath)^{n+1} \Psi_{n+1}(\rho)$.

We now apply this result to the evolution of any wave function over a quarter period. As for Eq.~(\ref{eqn:eexpn}), expand the wave function as a sum over the energy eigenfunctions: $\Psi (\xi)=\sum_{n=0}^{\infty} C_{n}\Psi_n (\xi)$. Then the evolved wave function at time $t$ is
\begin{equation} 
\Psi(\xi,t)=\sum\nolimits_{n=0}^{\infty}\exp[-\imath\omega t(n+\frac{1}{2})]C_{n}\Psi_n (\xi).
\end{equation} 
If $t=T/4=\pi/2\omega$, the factor $\exp[-\imath \omega t(n+\frac{1}{2})]$ becomes $\exp(-\imath\pi/4)(-\imath)^n$ and therefore
\begin{equation} 
\Psi(\xi,T/4)=e^{-\imath\pi/4}\sum\nolimits_{n=0}^{\infty}(-\imath)^n C_{n}\Psi_n (\xi).
\end{equation} 
But the Fourier transform of $\Psi (\xi)$ is $\Phi (\rho)=\mathcal{F}\Psi (\xi)=\sum_{n=0}^{\infty} C_{n}\mathcal{F}\Psi_n (\xi)=\sum_{n=0}^{\infty} (-\imath)^nC_{n}\Psi_n (\rho)$. Thus $\Psi(\xi,T/4)$ is, apart from the phase shift of $\pi/4$, the same function of $\xi$ as $\mathcal{F}\Psi (\xi)$ is a function of $\rho$, agreeing with the result from the propagator in Section \ref{sec:prop}.


\section{Transforming a state to stable form} 
We show that changing the scale of a state cannot yield a stable state unless $\Delta_{xp}=0$; but after a suitable phase change, a stable form can always be reached. 

If the moments of $\psi(x)$ are $\bar\Delta_x$, $\bar\Delta_{xp}$, $\bar\Delta_p$, and we change the scale of $\psi(x)$ to give $\chi(x)=\psi(s x)$, where $s$ is real, then it follows from the integrals for the expectation values that the moments of $\chi(x)$ are $\Delta_x^2=s^{-2}\bar\Delta_x$ and $\Delta_p^2=s^2\bar\Delta_p$, while $\Delta_{xp}$, $\Delta_x\Delta_p$ and $K$ do not change. Then $\epsilon$ becomes $\frac{1}{2}s^{-2}\bar\Delta_x^2/\alpha^2 +\frac{1}{2}
s^2\alpha^2\bar\Delta_p^2/\hbar^2$ and this has a minimum value of $\bar\Delta_x\bar\Delta_p/\hbar=(K^2+\bar\Delta_{xp}^2/\hbar^2)^{1/2}$ at $s^2=\hbar\bar\Delta_x/\alpha^2\bar\Delta_p$. The lowest energy depends on $\bar\Delta_{xp}$ and only if $\bar\Delta_{xp}=0$ can we reach $\epsilon=K$, which implies $A=0$.

We can change $\Delta_{xp}$ by a change in phase. Direct calculation gives the moments of $\phi(x)=\exp(-\imath x^2/2b^2)\psi(sx)$:
\begin{eqnarray}\label{eqn:Del_phi}
\Delta_x^2 &=& \bar\Delta_x^2/s^2, \\
\Delta_{xp} &=& \bar\Delta_{xp}-\hbar\bar\Delta_x^2/s^2 b^2, \\
\Delta_p^2 &=& s^2\bar\Delta_p^2-2\hbar\frac{}{}\bar\Delta_{xp}/b^2+\hbar^2\bar\Delta_x^2/s^2b^4.
\end{eqnarray}
It follows that $K$ is unchanged and that we can make $\Delta_{xp}=0$ by taking $b^2=\hbar\bar\Delta_x^2/s^2\bar\Delta_{xp}$. Then $\epsilon$ becomes $\frac{1}{2}(s^2\alpha^2K^2/\bar\Delta_x^2+\bar\Delta_x^2/s^2\alpha^2)$ and takes its minimum value $\epsilon=K$ with $s=\bar\Delta_x/\alpha\sqrt{K}$. Hence $b^2=\hbar\alpha^2K/\bar\Delta_{xp}$ and 
\begin{equation}\label{eqn:psi2phiA}
\phi(x)=\exp(-\frac{\imath}{\hbar}\frac{\bar\Delta_{xp}\,x^2}{2\alpha^2K})\psi(\frac{\bar\Delta_x\,x}{\alpha\sqrt{K}}).
\end{equation}
This wave function $\phi(x)$ is the stable equivalent of $\psi(x)$.

\section{Properties of real wave functions}
Many textbook examples have initial wave functions that are real, and therefore have special properties. These properties are ephemeral because every real wave function will immediately become complex (due to the term $\imath \hbar \,\partial_t \psi$ in Schr\"odinger's equation). However some special properties of real wave functions are useful in the present context. 

First we show that $\langle\hat p\rangle=0$ when $\psi$ is real. We use the Hermitean property of $\hat p=-\imath\hbar\partial_x$: $\int\!\phi^*\hat p\chi\,dx=\int\!(\hat p\phi)^*\chi\,dx$ for any two wave functions $\phi$ and $\chi$. Hence $\int\!\psi^*\hat p\psi\,dx=\int(\hat p\psi)^*\psi\,dx=\int(\imath\partial_x\psi)\psi\,dx\!=\!-\int\!\psi\,\hat p\psi\,dx$, since $\psi^*\!\!\!=\!\psi$. Similarly, $\Delta_{xp}=0$: $\int\!\psi\,\hat p\hat x\psi\,dx=\int(\hat p\psi)^*\hat x\psi\,dx=-\int(\hat p\psi)(\hat x\psi)\,dx=-\int\!\psi\,\hat x\hat p\psi\,dx$.

Unless $\langle\hat x\rangle=0$ also (in which case the centroid sits permanently at the origin), reality of the wave function implies that the centroid is instantaneously at rest at the extreme of its travel while $\Delta_x$ and $\Delta_p$ also take their extreme values. Thus the phases of the oscillations of the centroid are aligned with those of the second moments.

We can add momentum to any state $\phi$ using wave functions of the form $\psi=\exp(\imath p' x/\hbar)\phi$. In this case, $\langle\hat x\rangle_\psi=\langle\hat x\rangle_\phi$ while $\hat p \psi=p'\psi+e^{\imath p'x/\hbar}\hat p\phi$ and therefore $\langle\hat p\rangle_\psi=p'+\langle\hat p\rangle_\phi$. Also $\langle\hat x\hat p\rangle_\psi=p'\langle\hat x\rangle_\psi+\langle\hat x\hat p\rangle_\phi$ while $\langle\hat p\hat x\rangle=\langle\hat x\hat p\rangle-\imath\hbar$ for either $\psi$ or $\phi$, and therefore $\langle\hat p\hat x+\hat x\hat p\rangle_\psi=\langle\hat p\hat x+\hat x\hat p\rangle_\phi+2p'\langle\hat x\rangle_\phi$. If $\phi$ is real, it follows that $(\Delta_{xp})_\psi=0$.

\end{document}